\documentclass[11pt]{article}

\def\stacksymbols #1#2#3#4{\def\theguybelow{#2}
\def\verticalposition{\lower#3pt}
\def\spacingwithinsymbol{\baselineskip0pt\lineskip#4pt}
\mathrel{\mathpalette\intermediary#1}}
\def\intermediary#1#2{\verticalposition\vbox{\spacingwithinsymbol
\everycr={}\tabskip0pt
\halign{$\mathsurround0pt#1\hfil##\hfil$\crcr#2\crcr
\theguybelow\crcr}}}

\def\gapproxeq{\stacksymbols{>}{\sim}{4}{1}}
\def\lapproxeq{\stacksymbols{<}{\sim}{4}{1}}
\begin{document}
\begin{titlepage} 
\begin{center}
             \hfill    CfPA 00-th-01 \\
            \hfill    LPT-Orsay-00/68 \\
            \hfill astro-ph/0007452\\
\vskip .3in

{\large \bf Probing Large Distance Higher-Dimensional Gravity with Cosmic 
Microwave Background Measurements}\\[.1in]

Pierre Bin\'etruy$^1$ and Joseph Silk$^2$

{\em $^1$ Laboratoire de Physique Th\'eorique, Universit\'e Paris-Sud, 
Bat. 210

F-91405 Orsay Cedex, France

$^2$ Department of Astrophysics, University of Oxford, NAPL

Keble Road, Oxford OX13RH, United Kingdom}

\vskip .2in
\begin{abstract}
It has been recently argued that higher dimensional gravity theories may 
manifest themselves not only at short microscopic distances but also at large 
cosmological scales. We study the constraints that cosmic microwave background 
measurements set on such large distance modifications of the gravitational 
potential.
\end{abstract}
\end{center}
%\newpage
\pagestyle{empty}
\null
\end{titlepage}
\newpage
\renewcommand{\thepage}{\arabic{page}}
\setcounter{page}{1}
\def\thefootnote{\arabic{footnote}}
\setcounter{footnote}{0}

%\section{Introduction}

The possibility that the Universe may be higher-dimensional with standard 
matter and radiation localized on a four-dimensional surface known as 
a (3-)brane 
has attracted a lot of attention recently. This obviously would have 
important cosmological as well as astrophysical consequences: only the tip of 
this iceberg has  been explored.

In most of the models constructed until now, the laws of gravitation between 
two test masses on a 4-dimensional brane are standard 4-dimensional laws, for 
distances larger than a typical scale of order the radius of the 
compact dimensions \cite{ADD}.
When   one probes shorter distances, the full dimensions open up and 
the laws of gravitation become higher-dimensional.

Some of these extra dimensions may not even be compact. In the so-called 
Randall-Sundrum scenario \cite{prec,RaSu}, even though a fifth dimension is  
non-compact, there is a normalisable massless mode among the 5-dimensional
metric fluctuation modes, which is interpreted as the standard 4-dimensional
graviton. Strictly speaking, the fifth dimension is not infinite in the sense 
that there is an horizon at finite distance from the brane. Also, the bulk 
of spacetime has a simple anti-de Sitter structure.

The 4-dimensional matter feels long-distance gravity through its interaction 
with the 
5-dimensional bulk. In the case where this bulk has not such a simple 
structure as in the Randall-Sundrum model, one may expect  that the larger
the distance is, the more sensitive one becomes to the structure of the bulk, 
to the other branes which  it may contain
and thus to higher dimensions. This could have the effect that, at very large 
(cosmological?) 
distances, one may recover higher-dimensional gravity or at least be sensitive 
to scalar exchange  gravity, given the presence of moduli fields  describing 
compact dimension radii, or distances between branes in the bulk.

For example, if the large-distance physics associated with the cosmological 
constant is to be stabilized through some short-distance cancellations ensured 
by some bulk supersymmetry \cite{verlinde}, again one may expect that probing 
large distances 
on the brane may reveal violations of standard 4-dimensional gravitational 
laws.

Several models \cite{mill,GRS} which have been proposed recently show the 
type of behaviour  that we consider here.

In the GRS model \cite{GRS}, the (positive tension) 3-brane is located in the 
middle of a anti-de Sitter slab limited by two
negative tension branes and flat Minkowski  space on either  side. 
The solution 
considered for the metric includes a warp factor of the Randall-Sundrum type.
However, there are no normalisable zero mode. Instead, for an intermediate 
range of distances between two test masses on the 3-brane, the exchange of 
the collection of  non-normalisable graviton modes mimics the exchange of a 
single 4-dimensional graviton. This graviton is interpreted as a metastable 
state: when the distance becomes too large, it decays and one expects to 
recover 5-dimensional gravity. 

There is however a debate over the question whether the GRS model is 
internally consistent \cite{DGP}-\cite{PRZ}. The presence of a negative 
tension brane violates 
the weakest form of a positive energy  condition \cite{witten}.
Moreover \cite{GRS'},  the exchange of the radion scalar field (whose
vacuum expectation value fixes the interbrane distance --the radius--)
generates scalar antigravity which dominates at very large distances.  
This antigravity is clearly associated with the presence of  negative tension 
branes. It remains to be seen whether it is unavoidable \cite{CEH} and if 
the model remains  inconsistent at the quantum level. This source of
instability might also be cancelled by other scalar field exchanges, in
which case one would recover at large distances the 5-dimensional
gravity behavior.

%Moreover, it was shown \cite{GRS'} that concurrently to the effect that ensures
%4-dimensional gravity at intermediate scales, there is an one that generates
%scalar antigravity at very large distances. The nature of this scalar field is
%not clear \cite{GRS',PRZ}.  
%This antigravity is clearly associated with the presence of a negative tension 
%brane. It remains to be seen whether it is unavoidable \cite{CEH} and if it is 
%the sign of an inconsistency of the model.

In the models of Kogan {\em et al.} \cite{mill}, the dimensions are 
compact and exotic large distance effects are due to the presence of very 
light Kaluza-Klein states. It was however argued  \cite{KoRo} 
that this type of model and the preceding one  belong to the same general 
class. These models also 
have the possible drawbacks associated with negative tension branes.

Finally, it has been stressed recently \cite{DRT} that, when one tries
to give a small mass to localized scalars, the zero modes turn into
quasi-localized states with finite decay width: their exchange generates
a potential with a power law behavior at large distances.

In what follows, we will not rest on any specific model but consider 
the case where gravity becomes five-dimensional 
%ree different types of  modification of gravity 
at very large distances, only
commenting briefly on other possibilities.
How large  must such distances be? Milgrom \cite{MIL} has extensively discussed
deviations from Einstein gravity
on galactic scales in order to account for  galaxy rotation curves.
However because  dark matter is probed in virialised systems over a wide
range of density
on  scales from kpc to tens of Mpc, it
is clear that a simple change in the force law, such as we consider here,
could only occur on substantially larger scales. 
Hence if such violations appear at any macroscopic distances other than 
cosmological, existing limits  severely constrain them.
In the case of violations at cosmological distances, we argue in this note 
that cosmological background and deep galaxy redshift survey measurements may provide useful limits or 
interesting ways of probing such theories.

%\centerline{\bf CMB limits on higher dimensional gravity theories}

We recall that
in standard gravity, curvature or metric fluctuations scale as
$$ \delta \varphi = \frac{G\delta M}{rc^2} = \left( \frac{\delta M}{M} \right)
\left( \frac{r}{ct} \right)^2.$$
On subhorizon scales, the  
linear mass fluctuations in a gaussian-distributed density field
\cite{BBKS}
are
\begin{eqnarray}
\frac{\delta M}{M} &\approx& 0.1 \left( \frac{M_{eq}}{M} \right)
^{\frac{n+3}{6}} \qquad (M > M_{eq})\\
&\approx& {\rm constant} \ M^{\frac{n-1}{6}} \qquad (M \ll M_{eq})
\end{eqnarray}
where $M_{eq}$ is the horizon scale at the matter-radiation equal density
epoch and the density fluctuation
power spectrum has been taken to be $P(k) \propto k^n$. CMB measurements
confirm approximately
scale-invariant $(n \approx 1 )$ fluctuations $n = 1.2 \pm 0.3$ \cite{COBE} and the
power spectrum normalization on scales near $M_{eq}$ may be deduced from the
height of the first
acoustic peak \cite{BOOM}, to within an uncertainty of at most a factor of 2,
corresponding to the bias between
mass fluctuations (measured by the CMB, but dependent on cosmological model
parameters) and the fluctuations
in the luminous mass density (inferred from large-scale structure
surveys) \cite {GAW}.

If the gravity force changes on scale $r_s$ to a 5-dimensional law,
 the metric fluctuations can
be expanded as
$$ \delta \phi = \left( \frac{\delta M}{M} \right) \left( \frac{r}{ct}
\right)^2 \left( \frac{r_s}{r+r_s} \right).$$
Hence a scale-invariant fluctuation spectrum, as predicted by most
inflation and defect models for the fluctuations, results in large-scale
power with variance $\delta M / M \propto M^{- 1/3}$ as opposed to the
$M^{-2/3}$  predicted  for the standard gravity model
on scales $r<<r_s.$ 
This would not be visible 
as primary fluctuations on the last scattering surface of the 
CMB if $r_s$ is of order the horizon scale, since the Sachs-Wolfe effect
is generated by the constant potential fluctuations on horizon scales.
However it should give a signal that is potentially measureable in deep
redshift surveys such as 2DF and SDSS, which may eventually probe to 500
Mpc with galaxies, and can, using quasars, potentially probe much large
scales.

However the CMB fluctuations do provide a possible constraint on the
scale of higher dimensional gravity via the integrated Sachs-Wolfe
effect. This measures $\int \delta \dot\varphi c dt$ since last scattering.
The linear fluctuation growth rate is modified above scale $r_s$. To
demonstrate this, consider the Newtonian limit for the perturbation
equations as a simple approximation valid in the matter-dominated era on
subhorizon scales. One has
$$\left( \frac{\partial^2}{\partial t^2} + \frac{2\dot a}{a}
\frac{\partial}{\partial t} 
\right)\delta_k +
\frac{k^2}{a^2} \left( \frac{dp}{d \rho} - \varphi_k \right)\delta_k=0,$$
where $a(t)$ is the cosmological scale factor and $\rho ( \propto a^{-
3})$ is the density of non-relativistic matter. 
5-dimensional
gravity in the dust limit
modifies the potential in the small-scale limit  to  
\begin{eqnarray}
a^{-2} k^2 \varphi^{(5)}_k 
 &=& 4 \pi G \rho ( k/k_s),
\end{eqnarray}
and
a power-law solution $\delta_k \propto t^n$ satisfies
$$ n^2 + n/3 - (2/3)(k/k_s) = 0.$$

Hence growth is suppressed on scales $r > r_s$, and the usual
Jeans length is
modified to
$$ \frac{k_J}{a} = \frac{4 \pi G \rho}{dp/d \rho} \frac{a}{k_s} \qquad {\rm
or} \ \ \  L_J = \frac{2 \pi a}{k_J} =
r^{-1}_{s} v^2_s / (2 G \rho),$$
where $v^2_s = dp/d \rho$. The peak in the power spectrum of primordial
fluctuations is at $r_{eq}$, but 
the shape is slightly modified on larger scales 
\begin{equation}
r_{eq} \lapproxeq r \lapproxeq r_{h,ls} \label{bound}
\end{equation}
because of the suppression in subhorizon growth prior to last scattering.
On scales larger than the last scattering horizon $r_{h,ls}$, only the
potential fluctuations contribute to $\delta T/ T$. The change in growth
rate affects the shape of $P(k)$ near the peak, where power will be
suppressed.
 The contribution of the
integrated Sachs-Wolfe effect is also reduced.

More detailed numerical calculations are needed to produce precise
numbers, but it seems clear that the current precision of CMB
measurements on the scale range where growth modification is most
important already constrains $r_s \gapproxeq r_{h,ls}$.
These effects will not be easy to disentangle from the data
because of the effects of
cosmic variance, but a combination of redshift survey and CMB data
should be able to set significant limits on $r_s$.

One may also wonder whether the range of scale (\ref{bound}) may be
selected from the point of view of the fundamental theory. Let us note
here only that such scales are within a few orders of magnitude of the
length scale associated with the cosmological constant \cite{cosmocst}
and may therefore be associated with another cosmic coincidence
\cite{AHKM}. Indeed, the type of theories that we consider here have
been advocated for a partial solution of the cosmological constant problem.

\newpage
{\bf Acknowledgments}
\vskip .3cm
We wish to thank the hospitality of the Theory Group in Lawrence Berkeley 
National Laboratory (P.B.) and of the Center for Particle Astrophysics (J.S.) 
at U.C. Berkeley where part of this work  was done. 
One of us (J.S.) also thanks Martin Kunz for helpful comments.

\end{document}